\documentclass[showpacs,aps,twocolumn]{revtex4}
\usepackage{calc}
\usepackage{graphicx}
\usepackage{dcolumn}
\usepackage{bm}
\usepackage{amsmath}

\begin{document}

\title{Delocalization and wave-packet dynamics in one-dimensional diluted Anderson
models}
\author{F.~A.~B.~F. de Moura$^a$,  M.~N.~B.~Santos$^b$,  U.~L.~Fulco$^b$,
M.~L.~Lyra$^a$, E.~Lazo$^c$, M.~E.~ Onell$^c$}

\affiliation{(a)Departamento de F\'{\i}sica, Universidade Federal
de Alagoas, 57072-970 Macei\'o, AL, Brazil}

\affiliation{(b) Departamento de F\'{\i}sica, Universidade Federal do Piaui,
64048550 Teresina, PI, Brazil}

\affiliation{(c)Departamento de
F\'{i}sica, Fac. de Ciencias, U. de Tarapac\'{a}, Casilla 7-D,
Arica, Chile}

\begin{abstract}

We study the nature of one-electron eigen-states in a
one-dimensional diluted Anderson model where every Anderson
impurity is diluted by a periodic function $f(l)$ . Using
renormalization group and transfer matrix techniques, we provide
accurate estimates of the extended states which appear in this
model, whose number depends on the symmetry of the diluting
function $f(l)$. The density of states (DOS) for this model is
also numerically obtained and its main features are related to the
symmetries of the diluting function $f(l)$. Further, we show that
the emergence of extended states promotes a sub-diffusive spread
of an initially localized wave-packet.

\end{abstract}


\maketitle

\section{Introduction}

The nature of electronic states in disordered tight-binding models with
site-diagonal uncorrelated disorder was firstly studied by Anderson~\cite
{anderson} when the localization of quantum states was discussed in
connection with the transport properties of a random lattice. One of the
most remarkable effects of disorder, which has been demonstrated by several
authors, is the exponential localization of all one-electron eigen-states in
the one-dimensional Anderson model, irrespective to the strength of disorder~
\cite{abrahams}. The electron localization in random lattices is a result of
destructive quantum interference due to incoherent backscattering.
Therefore, the localization phenomenon occurs in general model systems
involving wave propagation in random media.

Recently, the existence of delocalized states in several variants
of the low-dimensional Anderson models has been reported~
\cite{flores1,dunlap,sen,datta,evangelou1,adame1,adame2,flores2,ojeda,chico,izrailev,xiong1,cervero,liu1,ref1,ref2,massa}.
In these works, the presence of short or long-range correlations
appears as the fundamental mechanism responsible for the emergence
of extended states. This theoretical prediction about suppression
of localization was recently confirmed experimentally in doped
polyaniline~\cite{Wu}, semiconductor superlattices with
intentional correlated disorder~\cite {bellani} as well as
microwave transmission spectra of single-mode waveguides with
inserted correlated scatterers~\cite{apl2}. Among these models,
the diluted Anderson chain has attracted a renewed interest~\cite
{Hilke,adame,lazo1,deng,lazo2}. Hilke ~\cite{Hilke} introduced an
Anderson model with diagonal disorder diluted by an underlying
periodicity. The model consisted of two interpenetrating
sub-lattices, one composed of random potentials (Anderson lattice)
and the other composed of non-random segments of constant
potentials. Due to the periodicity, special resonance energies
appear which are related to the lattice constant of the non-random
lattice. The number of resonance energies is independent of the
system size and, therefore, it was conjectured that these states
shouldn't have any influence on the transport properties in the
infinite size limit. In ref.~\cite{adame} the authors presented a
simple model for alloys of compound semiconductors by introducing
a one-dimensional binary random system where impurities are placed
in one sublattice while host atoms lie on the other sublattice.
The existence of an extended state at the band center was
demonstrated, both analytic and numerically. The diluted Anderson
model was recently extended to include a general diluting function
which defines the on-site energies within each non-random
segment~\cite{lazo1}. Using a block decimation approach, it was
demonstrated that this model displays a set of extended states,
the number of which strongly depends on the length of the diluting
segments and the symmetries of the diluting function.

In this work we will use the matrix decimation method and the transfer
matrix technique to provide accurate estimates of the set of extended states
in general 1D diluted Anderson models. The density of states (DOS) will be
shown to display a set of gaps whose number and width depend on the set of
impurities used to dilute the disordered lattice. Further, we will
investigate the influence of such resonant extended states on the
wave-packet dynamics. Our results suggest that, in spite of the number of
resonance energies being independent of the system size~\cite{Hilke}, these
extended states induce a sub-diffusive spread of an initially localized
wave-packet.

\section{Model and Formalism}

The standard one-dimensional Anderson model is described by a tridiagonal
Hamiltonian
\begin{equation}
H=\sum_{j}\epsilon _{j}|j><j|+t\sum_{j}[|j><j+1|+|j><j-1|]
\end{equation}
where disorder is introduced on the site energies $\epsilon _{j}$ which are
uncorrelated random numbers chosen from a previously defined distribution.
In our calculations, we will use energy units such that the hopping term $t=1
$ and the random site energies will be taken uniformly from the interval $%
[-0.5,0.5]$. The diluted Anderson model is constructed by introducing a sequence
of $k$ new sites between each original pair of neighboring sites. These
sequences are all identical and the on-site energies within such sequences
are given by $f(l)$, $l=1,2,...,k$.

To study the properties of the one-electron eigen-states, we employ a
transfer matrix calculation (TM) in order to obtain the Lyapunov exponent
defined as the inverse of the localization length $L_{c}$. The Schroedinger
equation for the present tight-binding model is:
\begin{equation}
\epsilon _{n}c_{n}+c_{n-1}+c_{n+1}=Ec_{n},
\end{equation}
where $\Phi =\sum_{n}c_{n}|n>$ is an eigenstate with energy $E$. The above
equation can be rewritten as a transfer matrix equation
\begin{equation}
\left(
\begin{array}{c}
c_{n+1} \\
c_{n}
\end{array}
\right) =\left(
\begin{array}{cc}
E-\epsilon _{n} & -1 \\
1 & 0
\end{array}
\right) \left(
\begin{array}{c}
c_{n} \\
c_{n-1}
\end{array}
\right) ,~n=0,1,2,...,N~.
\end{equation}
Based on the asymptotic behavior of the matrix product $\prod_{n=1}^{N}Q_{n}$%
, where $Q_{n}$ are $2\times 2$ transfer matrices, the Lyapunov exponent $%
\gamma $ can be defined as:
\begin{figure}[tbp]
\centering
\includegraphics[width=64mm,clip]{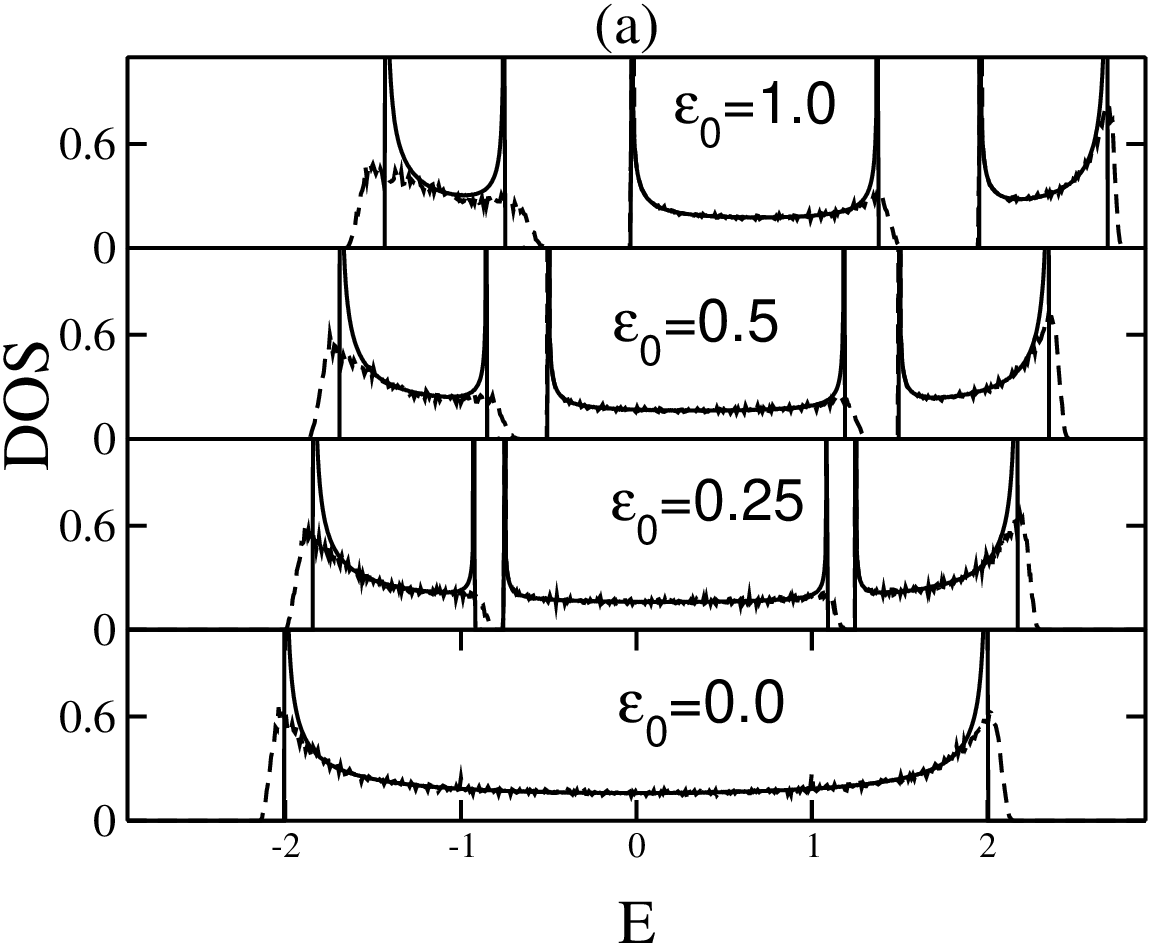} %
\includegraphics[width=64mm,clip]{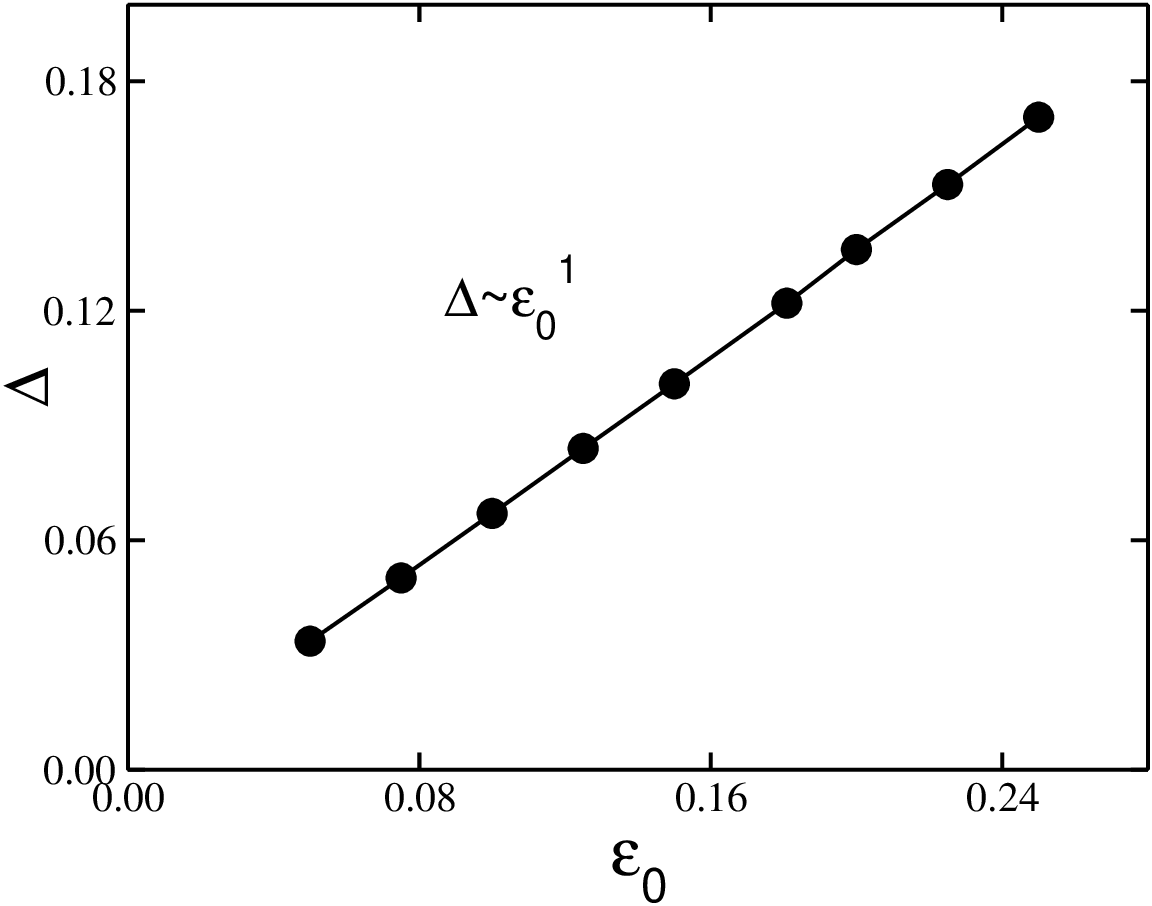}
\caption{(a) Dashed lines represent the Normalized density of
states (DOS) as a function of energy $E$ obtained using the Dean's
method for a chain with $N=10^{5}$ sites of the diluted 1D
Anderson model with dimer segments of identical on-site energies
$\protect\epsilon_0$. The DOS displays two pseudo-gaps for
$\protect\epsilon_0>0$, which are reminiscent of the gaps present
in the DOS of the corresponding pure models (shown as solid
lines). (b) The width of the gap ($\Delta $) for a diluted chain
without disorder versus $\protect\epsilon_0$. The width $\Delta $
displays a linear behavior $\Delta \propto \protect\epsilon_0^{1}$
for any length of the diluting segment.}
\end{figure}
\begin{figure}[tbp]
\centering
\includegraphics[width=64mm,clip]{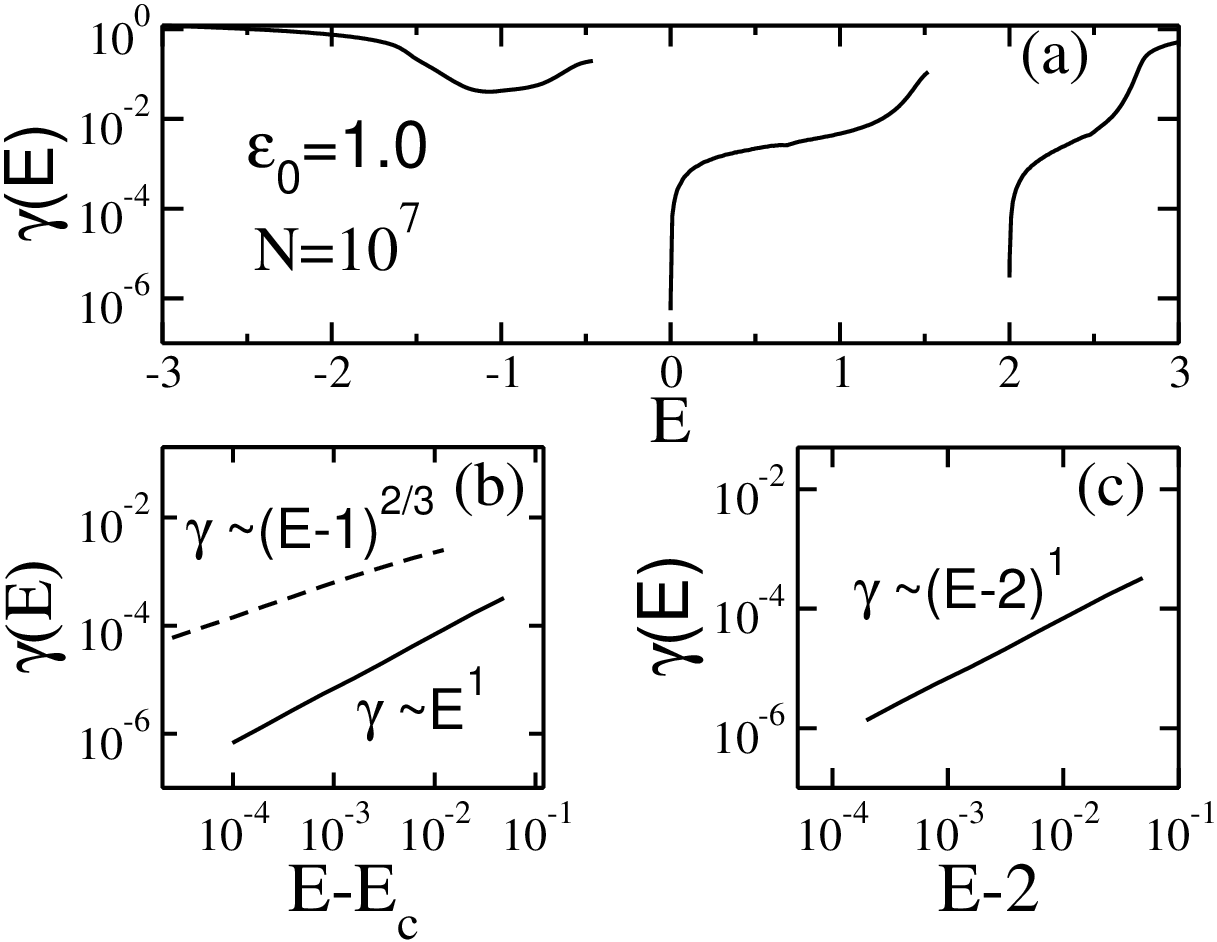} %
\includegraphics[width=64mm,clip]{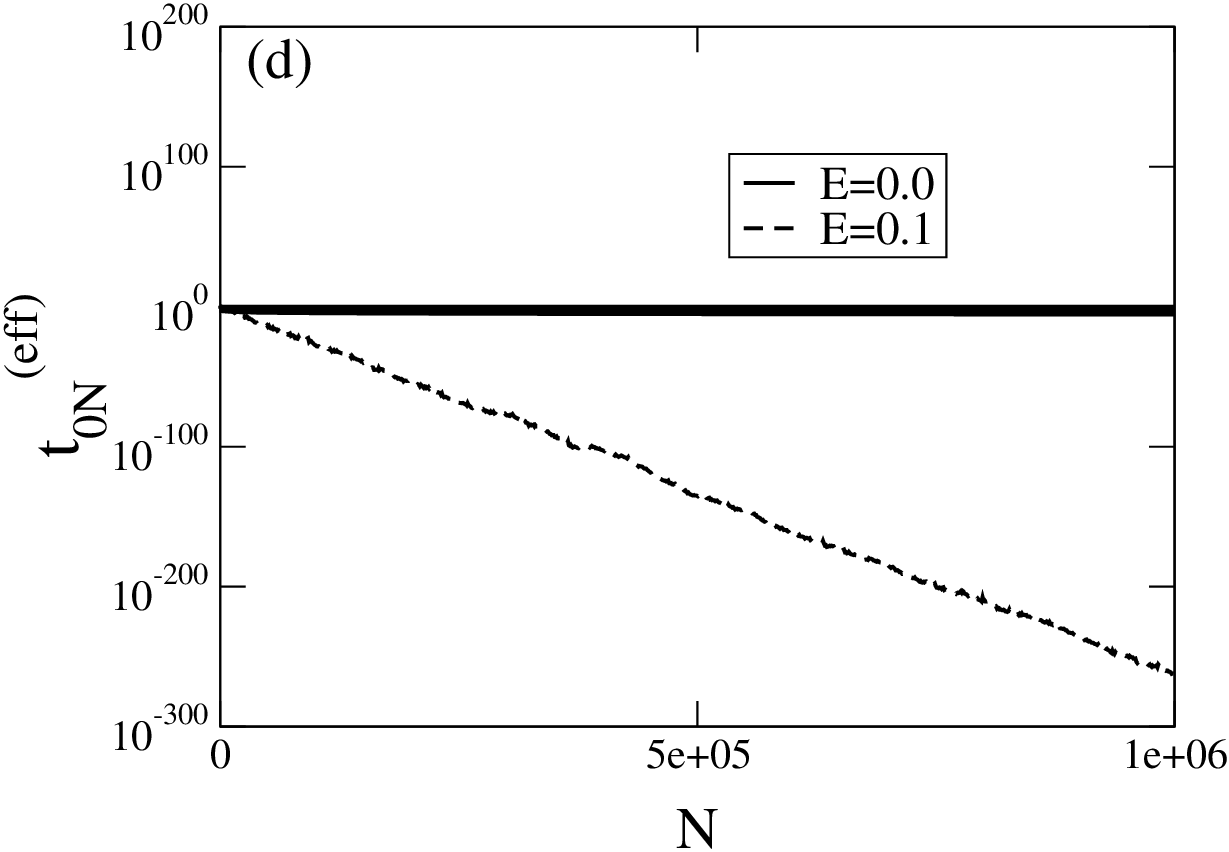}
\caption{a) Lyapunov exponent $\protect\gamma $ versus $E$ for a
chain with $ N=10^{7}$ sites diluted by dimers with
$\protect\epsilon _{0}=1.0$~. We found that $\protect\gamma $ is
finite for all energies except at the two resonance energies where
$\protect\gamma $ vanishes linearly (see b) and c)). For
$\protect\epsilon _{0}=0.0$, one has a slower non-linear vanishing
of $\protect\gamma \propto (E-E_c)^{(2/3)}$ (see long-dashed line
in fig b).
  (d) The
effective hopping $t_{0,N}^{(eff)}$ displays an oscillating
behavior as a function of $N$ for $E=0.0$~(extended state) and an
exponentially decaying behavior for $E=0.1$ (localized state). }
\end{figure}
\begin{equation}
\gamma =\lim_{N\rightarrow \infty }\frac{1}{N}\log \frac{%
|\prod_{n=1}^{N}Q_{n}z(0)|}{|z(0)|},
\end{equation}
where  $z(0)={u_1 \choose u_0}$ is a generic initial condition. In
addition, the Lyapunov exponent can be obtained using a decimation
renormalization-group (DRG) technique which is based on the
particular form
of the equation of motion satisfied by the Green's operator matrix elements $%
[G(E)]_{i,j}=\langle i|1/(E-H)|j\rangle $\cite{chico}:
\begin{eqnarray}
(E-\epsilon _{n+\mu }^{0})[G(E)]_{n+\mu ,n} &=&\delta _{n+\mu ,n}+ \\
&&t_{n+\mu ,n+\mu -1}^{0}[G(E)]_{n+\mu -1,n}+  \nonumber \\
&&t_{n+\mu ,n+\mu +1}^{0}[G(E)]_{n+\mu +1,n},  \nonumber
\end{eqnarray}
where $\epsilon _{n}^{0}=\epsilon _{n}$ and $t_{n,n+1}^{0}=t_{n-1,n}^{0}=t$.
After eliminating the matrix elements associated with a given site, the
remaining set of equations can be expressed in the same form as the original
one, but with energies and hopping amplitudes renormalized. Therefore, the
decimation-renormalization technique consists in removing iteratively the
sites $1,2,...,N-1$ of the system obtaining in that way the effective
energies of the external sites and the effective hopping between them by
using the iterative equations
\begin{eqnarray}
\epsilon _{N}^{(N-1)}(E) &=&\epsilon _{N}+t_{N-1,N}\frac{1}{E-\epsilon
_{N-1}^{(N-2)}(E)}t_{N-1,N}, \\
t_{0,N}^{(eff)}(E) &=&t_{0,N-1}^{(eff)}\frac{1}{E-\epsilon _{N-1}^{(N-2)}(E)}%
t_{N-1,N},
\end{eqnarray}
where, after $N-1$ decimations, $\epsilon _{N}^{N-1}$ denotes the
renormalized diagonal element at site $N$ and $t_{0,N}^{(eff)}$ indicates
the effective renormalized hopping connecting the sites $0$ and $N$. The
behavior of the effective interaction $t_{0,N}^{(eff)}$ during the
decimation procedure can be used as another evidence of the
localized/delocalized nature of the one-electron states. An oscillatory
behavior of $t_{0,N}^{(eff)}$, as a function of $N$, signals an extended
state. On the contrary, an exponentially localized state results in an
exponential decrease of $t_{0,N}^{(eff)}$ as the decimation process
proceeds. After a large number of decimation steps, the Lyapunov exponent is
asymptotically related to the effective hopping $t_{0,N}^{(eff)}$ in the
following way:
\[
\gamma (E)=-\lim_{N\rightarrow \infty }[\frac{1}{N}\ln
|t_{0,N}^{(eff)}(E)|].
\]

\section{Results and Discussion}

In the present study, we use chains with $10^{7}$ sites in order
to calculate the Lyapunov exponent within the entire energy band.
We start our analysis investigating the main features of a diluted
1D Anderson model with segments consisting of two sites with
identical on-site energies $\epsilon _{0}$. The emergence of
extended states for this particular dilution was analytically
demonstrated in Ref~\cite{Hilke,lazo1}. The density of states
(DOS) for general 1D tight-binding models can be obtained by using
the negative eigenvalue theorem\cite{dean}. In Fig.~1(a) we show
typical plots of the DOS of chains with $N=10^{5}$ sites and
several $\epsilon _{0}$ values. In this diluted chain, the density
of states displays two pseudo-gaps for $\epsilon _{0}>0$. The
widths of these pseudo-gaps increase as a function of $\epsilon
_{0}$. These pseudo-gaps are reminiscent of the gaps present in
the absence of disorder and their width is proportional to
$\epsilon _{0}$ for any length of the diluting segment (see
fig.~1b), thus following a similar trend of the gap found in the
two band model~\cite{adametwo}. The presence of disorder rounds
one of the band edges which displays an exponentially decaying
tail. On the other hand, the opposite band edge remains as a
singularity of the DOS. The surviving of this singularity in the
presence of disorder will be shown to influence the wave-packet
dynamics of the diluted Anderson model.

\begin{figure}[tbp]
\centering
\includegraphics[width=64mm,clip]{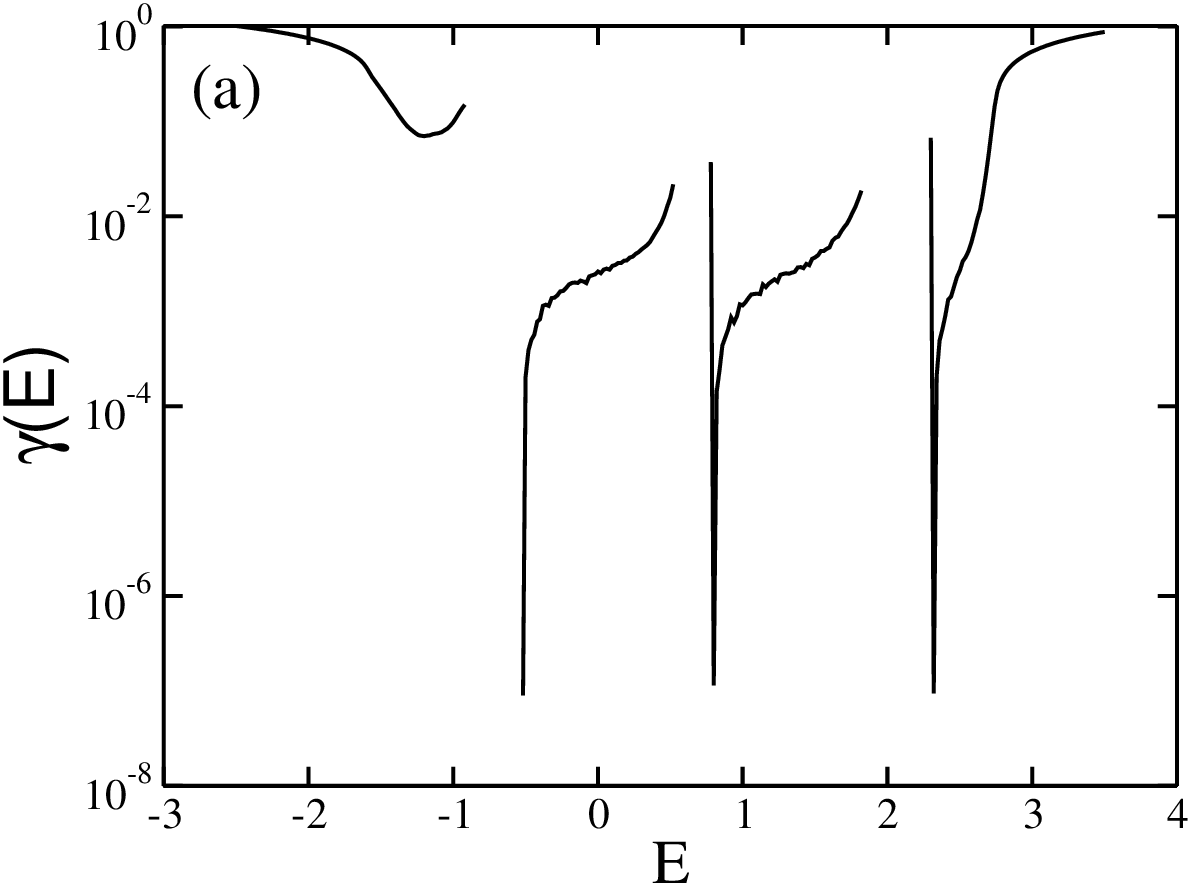} %
\includegraphics[width=64mm,clip]{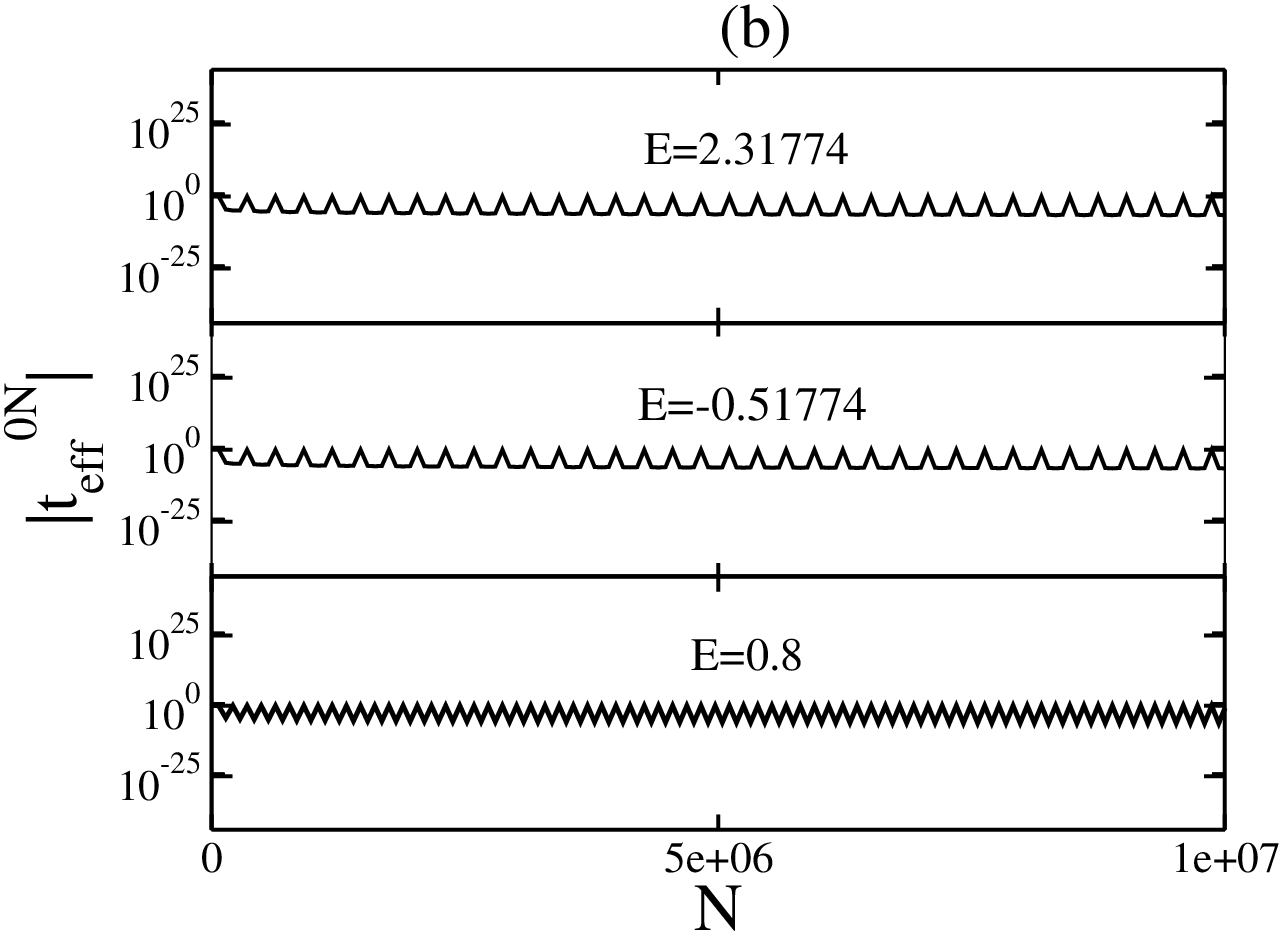}
\caption{a) The Lyapunov exponent $\protect\gamma$ versus $E$ for an
Anderson chain diluted by segments with $k=3$ sites exhibiting specular
symmetry $f(l)=\{\protect\epsilon_2,\protect\epsilon_3,\protect\epsilon_2\}$ where
$\protect\epsilon _{2}=0.8$ and $\protect\epsilon_{3}=1.0$.
All calculations were performed for a chain with $N=10^{7}$ sites. (b) The
effective hopping $t_{0,N}^{(eff)}$ as a function of $N$ for the resonance
energies $E_{c}=\{0.8,2.31774,-0.51774\}$ corresponding to extended states.}
\end{figure}
\begin{figure}[tbp]
\centering
\includegraphics[width=64mm,clip]{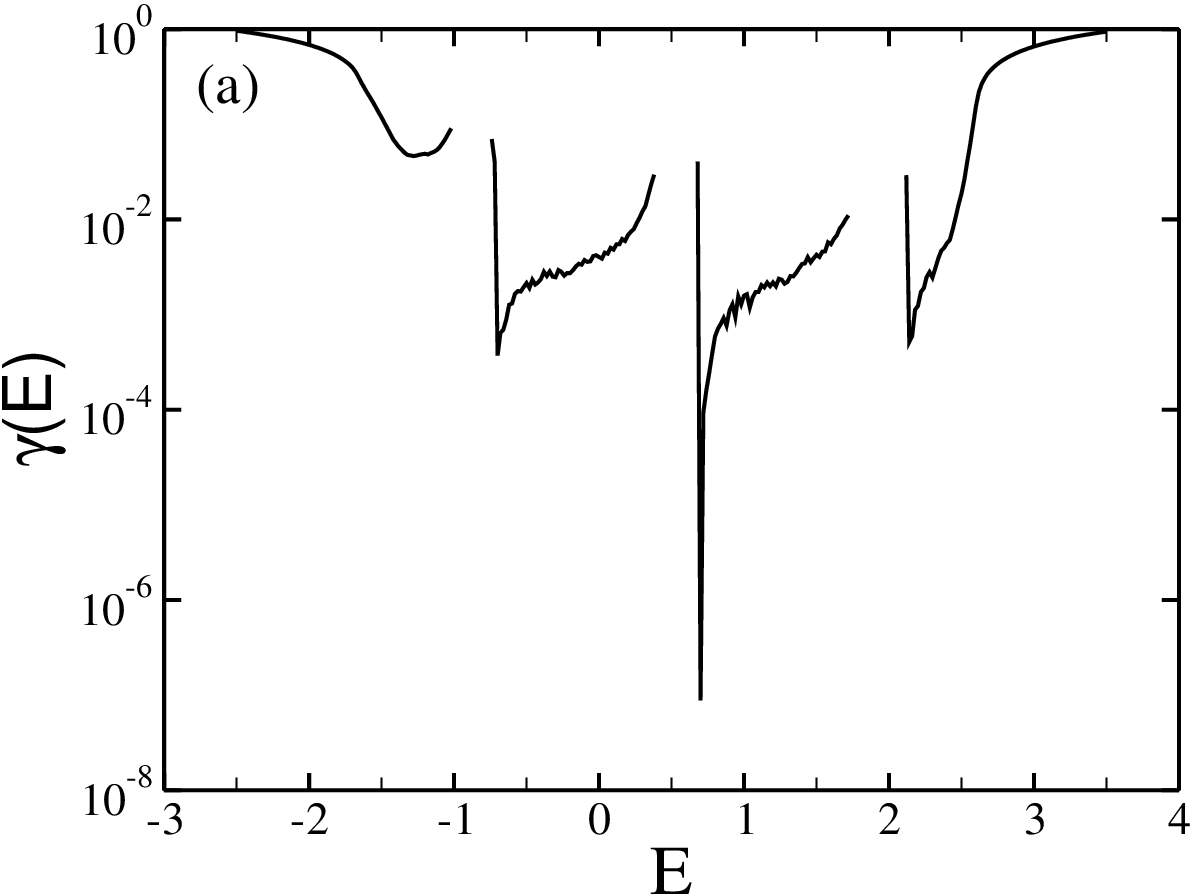} %
\includegraphics[width=64mm,clip]{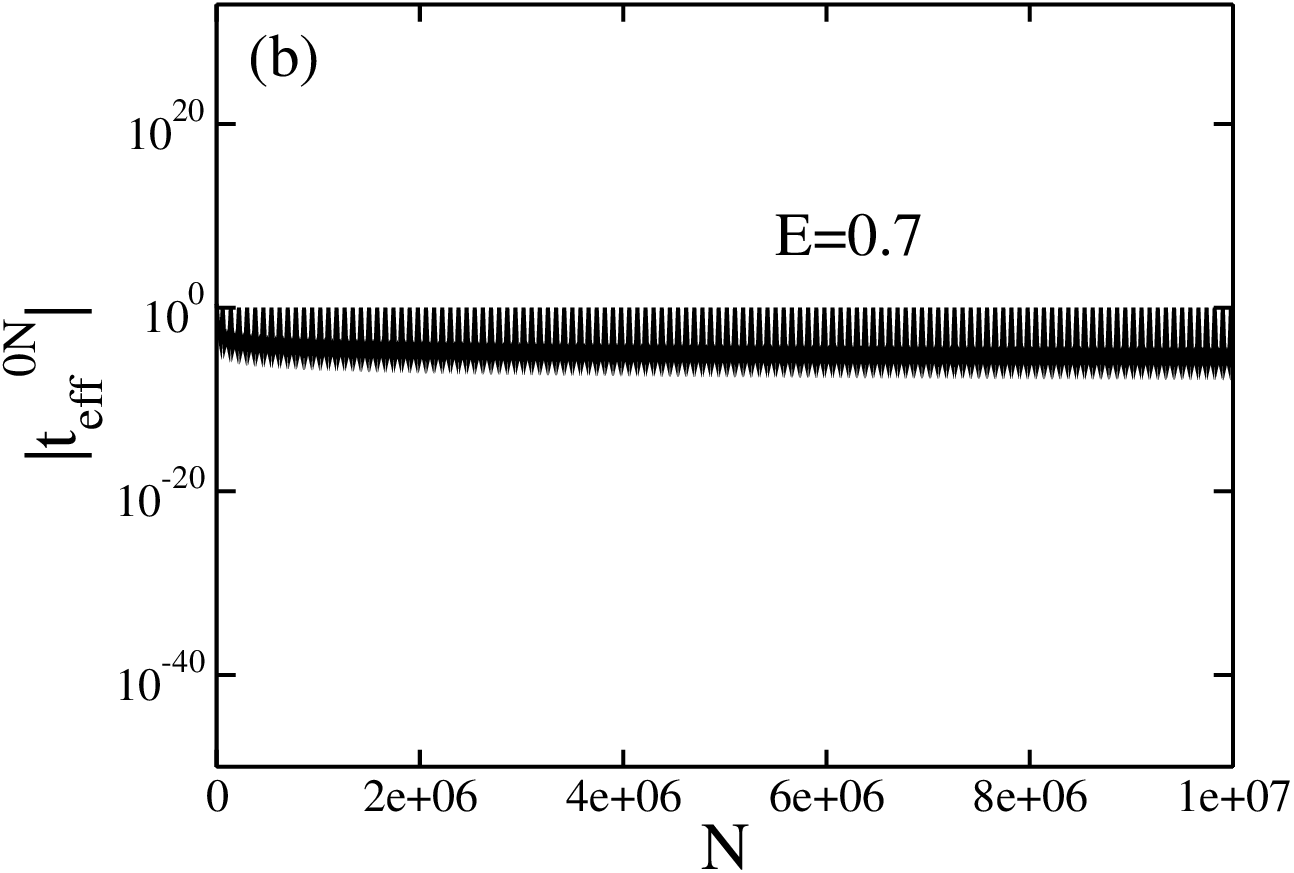}
\caption{(a) The Lyapunov exponent $\protect\gamma $ versus $E$ for an
Anderson chain with $N=10^{7}$ sites diluted by segments with $k=3$ sites
exhibiting the distance to center symmetry $f(l)=\protect\epsilon
_{l}=\left\{ 0.8,0.7,0.6\right\}$. In this case, there is a single extended
state located at the resonance energy $E=0.7.$ (b) The effective hopping $%
t_{0,N}^{(eff)}$ as a function of $N$ for $E=0.7$ shows the oscillating
behavior characteristic of extended states.}
\end{figure}

In Fig.~2(a) we depict the Lyapunov exponent $\gamma $ versus $E$
for chains with $N=10^{7}$ sites and $\epsilon _{0}=1.0$. We
calculate $\gamma $ using both transfer matrix and decimation
renormalization group approaches. The results were identical
within our numerical accuracy. We found that $\gamma (E)$ is
finite for all energies except at the two resonance energies
($E=0.0$ and $E=2.0$) where the Lyapunov exponent is of the order
$\frac{1}{N}$. Therefore, all states are exponentially localized
except at these two resonance energies where $\gamma $ vanishes in
the thermodynamic limit. This result coincides with the analytical
prediction of extended states at $ E=\epsilon _{0}\pm 1$
\cite{lazo1}. In Fig.~2(b) and 2(c), we point out the linear
vanishing of $\gamma $ around $E=0$ and $E=2.0$ (see full lines in
fig Fig.~2(b) and 2(c)). For $\protect\epsilon _{0}=0.0$, we found
a slower non-linear vanishing of $\protect\gamma \propto
(E-E_c)^{(2/3)}$ (see dashed line in fig.~2(b)), in perfect
agreement with ref.~\cite{Hilke}. The larger exponent found for
$\epsilon _0\neq 0$ can
 be attributed to fact that the resonance energies for $\epsilon_0\neq 0$ are precisely at
 the DOS band edge singularities, which are absent
for $\epsilon _0 = 0$.  To further characterize the extended
nature of these resonant states, we plot in Fig.~2d the effective
interaction $t_{0,N}^{(eff)}$ versus $N$. For the resonant state
at $E=0.0$ the Lyapunov exponent $\gamma $ vanishes due to the
oscillating behavior of $t_{0,N}^{(eff)}$. The extended character
of this state is, therefore, reflected by a finite effective
hopping amplitude between the sites located at the chain ends. For
$E=0.1$ the effective hopping decreases exponentially, indicating
the localized character of non-resonant states. We have also
considered longer diluting sequences with constant on-site energy.
The number of extended states always corresponds to the number of
sites in the sequence, as expected \cite{Hilke,lazo1}.

From the theoretical point of view, the basic condition to find
delocalized states in diluted disordered systems is the existence
of some symmetries in the periodic function $f(l)$ which defines
the site energies $\epsilon _{l}=f(l)$ within the diluting
segment. The relevant symmetries are\cite {lazo1}:\newline a)
Specular reflection symmetry with respect to center of the
diluting segment, namely, $f(l)=f(k+1-l)$, with
$l=1,2,...k$.\newline
b) Distance to the center symmetry, namely, $%
|f(l_{0})-f(l_{0}-j)|=|f(l_{0})-f(l_{0}+j)|$, with $j$ ranging from $j=1$ to
$j=\left(k-1\right)/2$ and $l_{0}$ being the position of the central segment
site. This kind of symmetry can be present only in segments with an odd
number of sites.

Delocalized states emerge whenever the diluting function exhibits
one of the above symmetries. In the following, we consider two
examples using diluting segments with either one of the above
symmetries present.

To observe the emergence of extended states in the diluted Anderson model
with specular reflection symmetry, we considered a diluting segment with $k=3
$ sites: $f(l)=\{\epsilon _{2},\epsilon _{3},\epsilon _{2}\},$ where the
site energies are $\epsilon _{2}=0.8$ and $\epsilon _{3}=1.0$, respectively.
In fig.~3(a) we show the Lyapunov exponent $\gamma $ versus $E.$ All
calculations were performed for $N=10^{7}$ sites. After executing the
renormalization process, we found that the Lyapunov exponent vanishes for $%
E_{c}\approx \left\{ 0.8,2.31774,-0.51774\right\} ,$ in agreement with
the block decimation result \cite{lazo1}.  In addition, we show in Fig.~3(b)
the oscillating pattern exhibited by the effective hopping $t_{0,N}^{(eff)}$
as a function of $N$ for the same critical energies, thus confirming the
extended nature of these states.

A similar analysis can be made to investigate a diluted Anderson chain with
the distance to the center symmetry. We considered, for this case, the
following diluting function: $f(l)=\epsilon _{l}=\left\{ 0.8,0.7,0.6\right\}
$. From the theoretical calculation of \cite{lazo1}, one can anticipate that
this chain will have only one extended state with energy $E_c=0.7$~.
Fig.~4(a) and Fig.~4(b) clearly show that the Lyapunov exponent vanishes
only for $E_c=0.7$ and the effective hopping $t_{0,N}^{(eff)}$ displays an
oscillating behavior at this critical energy.
\begin{figure}[ht]
\centering
\includegraphics[width=60mm,clip]{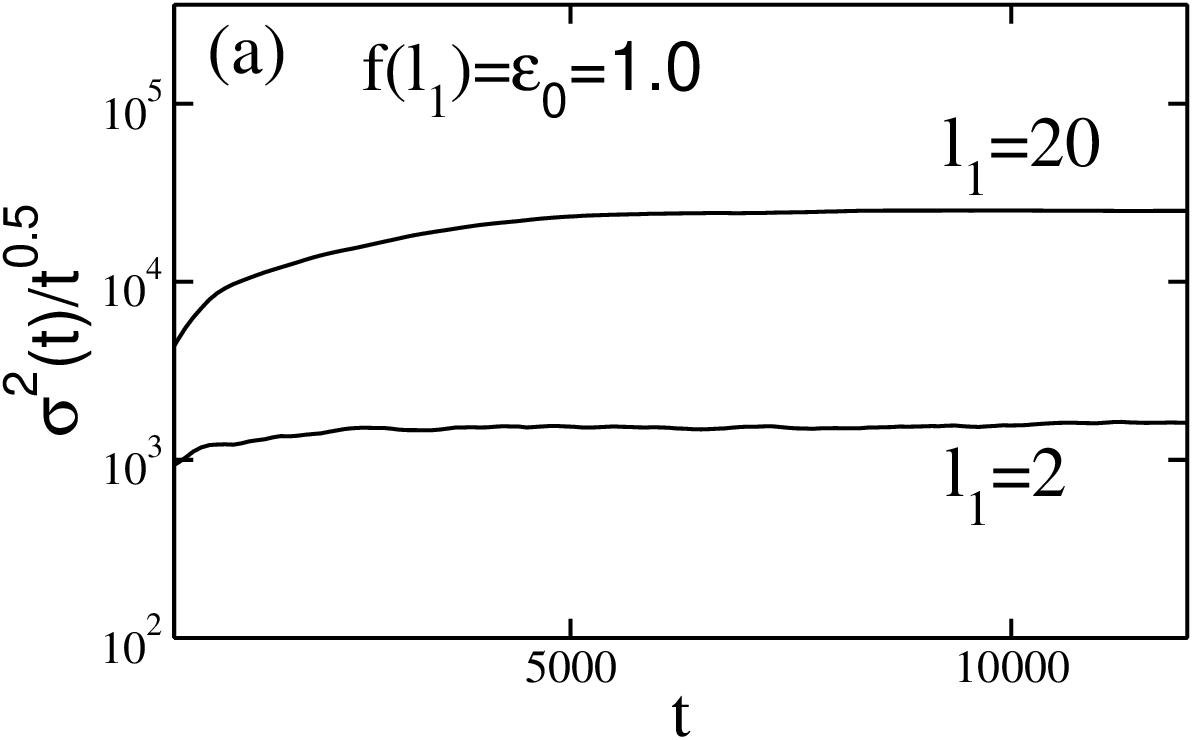}
\includegraphics[width=60mm,clip]{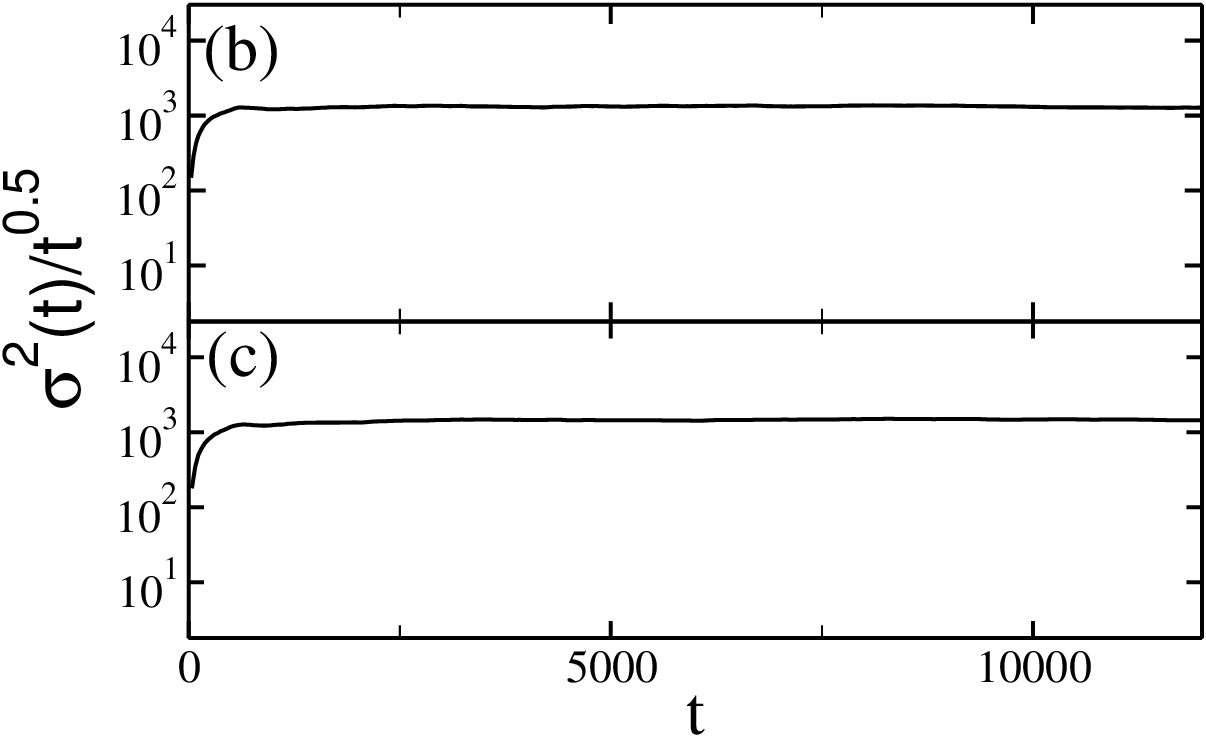}
\caption{ The scaled mean-square displacement $\protect\sigma
^{2}(t)/t^{0.5} $ versus times $t$ for the 1D diluted Anderson
model with $7\times 10^{4}$ sites for: a) Constant diluting
function $\protect\epsilon _{0}=f(l_{1})= \protect\epsilon _{0}=1$
with $l_{1}=2,20$. (b) Diluting function with
specular reflection symmetry $f(l)=\{\protect\epsilon _2,\protect\epsilon_3,\protect\epsilon_2\}$ with $\protect\epsilon_2=0.8$ and $\protect%
\epsilon_3=1.0$ (c) Diluting function with the distance to center
symmetry $f(l)=\{0.8,0.7,0.6\}$. A sub-diffusive behavior
$\protect\sigma ^{2}\propto t^{0.5}$ takes place for any symmetry
supporting extended states. }
\end{figure}
\begin{figure}[ht]
\centering
\includegraphics[width=64mm,clip]{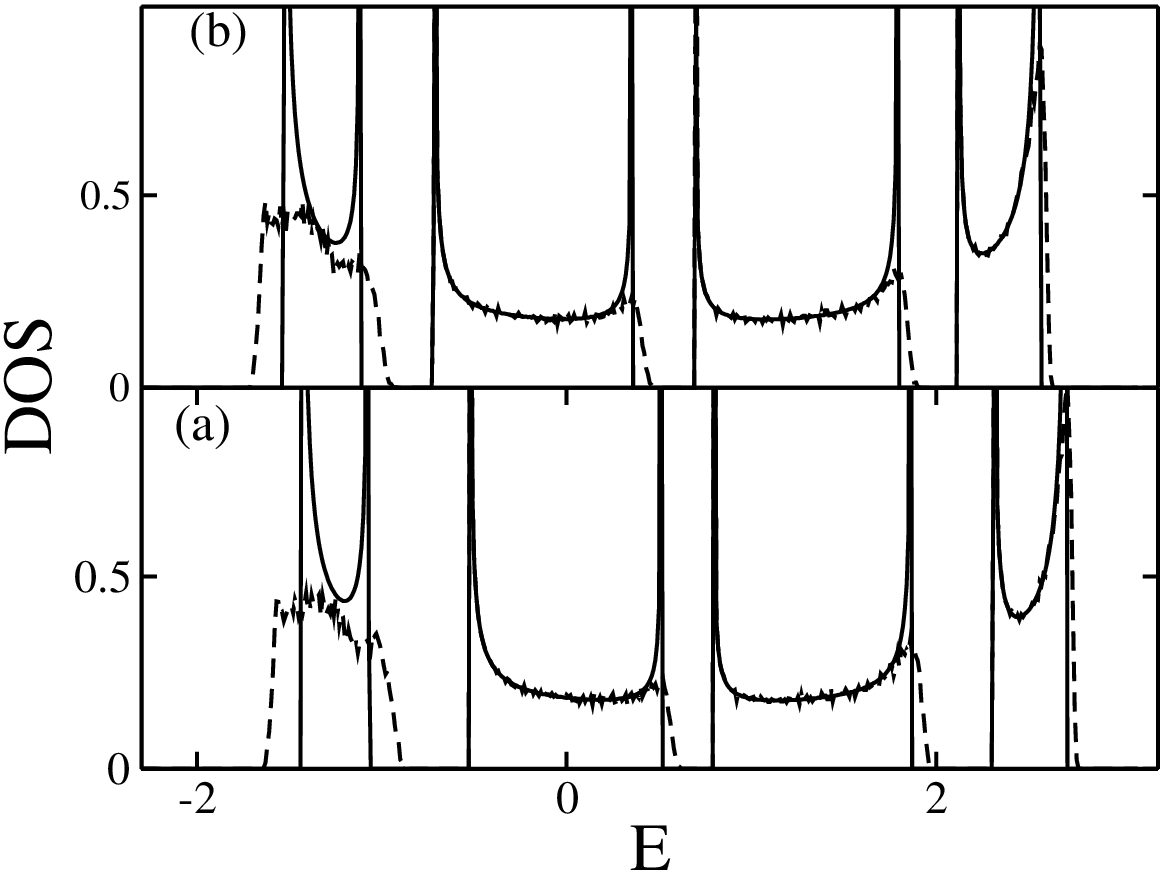}
\caption{Dashed lines represent the   normalized density of states (DOS) for the diluted chain with
a) specular reflection symmetry $f(l)=\{0.8,1.0,0.8\}$ and b) distance to
the center symmetry $f(l)=\{0.8,0.7,0.6\}$. In both cases, the resonance
energies associated with extended states are coincident with DOS
singularities.The DOS displays  pseudo-gaps for
$\protect\epsilon_0>0$, which are reminiscent of the gaps present
in the DOS of the corresponding pure models (shown as solid
lines).}
\end{figure}

In order to investigate the influence of the above resonant
extended states on the transport properties of the diluted
Anderson model, we calculate the time dependence of the
mean-square displacement of an initially localized wave-packet.
Starting with one electron fully localized at the  Anderson site
closer to the chain  center, the wave-function amplitudes
$b_{n}(t)$ were obtained from the following equations of motion:
\begin{equation}
i\dot{b}_{n}(t)=\epsilon _{n}b_{n}(t)+(b_{n-1}(t)+b_{n+1}(t))\hspace{0.5cm}%
n=1,2,...,N.  \label{Runge}
\end{equation}
Using a Runge-Kutta algorithm, we solve the above set of coupled equations
and calculate the mean-square displacement
\begin{equation}
\sigma ^{2}(t)=\sum_{n}(n-n_{0})^{2}|b_{n}(t)|^{2}.
\end{equation}
To minimize end effect, our numerical calculation was performed in
a very large chain with $N=7\times 10^{4}$ sites. As a
consequence, the site amplitudes at the ends of the chain
($b_{1}(t)$ and $b_{N}(t)$) are always negligible. The numerical
integration of equations (8) was performed using the Runge-Kutta
algorithm with precision $\Delta t$ smaller than $10^{-2}$. In
Fig.~5(a-c), we exhibit our main results for the time evolution of
the mean-square displacement $\sigma^{2}(t)$ on diluted chains
considering all relevant symmetries of the diluting function.  In
all cases, we obtained a clear sub-diffusive behavior: $\sigma
^{2}\propto t^{0.5}$. This sub-diffusive spread of the wave-packet
is related to the fact that the resonance energies supporting
extended states occur at band edge singularities of the DOS, as
depicted in Fig.~6.  A similar diffusive-like spread of the
wave-packet has also been observed to occur with collective
excitations in other disordered systems with the Lyapunov exponent
vanishing at DOS singularities as, for example, in random harmonic
chains~\cite {datta2} and disordered ferromagnetic
chains~\cite{evangelou1,mauricio}.

\section{Summary and Conclusions}

In summary, we investigated the 1D diluted Anderson model where
every Anderson impurity is diluted by a set of site energies given
by a diluting function $f(l)$. Using the renormalization group
approach and the transfer matrix technique, we obtained the
Lyapunov exponent $\gamma $ in several diluting cases exploring
distinct symmetries supporting extended states, such as constant
diluting function, diluting function with either specular
reflection or distance to the center symmetry. In all cases, we
identified the resonance energies at which the model exhibit
extended states, in full agreement with previous
results~\cite{Hilke,lazo1}. We also studied the temporal spread of
an initially localized electron wave-function in these diluted
chains by following the time dependence of the wave-packet
mean-square displacement. We found that, associated with the
presence of extended states at resonance energies located at DOS
band edge singularities, the electron wave-packet displays a
sub-diffusive spread($\sigma ^{2}(t)\propto t^{0.5}$). This result
is consistent with the general picture that the wave-packet
dynamics depends on the relative location of resonance energies
and DOS singularities~\cite{kundu1}.

\section{Acknowledgments}

It is a pleasure to acknowledge the hospitality of Prof. Dr. Arkady Krokhin
and Prof. Dr. Felix Izrailev in Puebla during the III International Workshop
on Disordered Systems where this work was initiated. This work was partially
supported by the Brazilian research agencies CNPq and CAPES, as well as by
the Alagoas state research agency FAPEAL. The support of this research by
the Direcci\'{o}n de Investigaci\'{o}n y Postgrado de la Universidad de
Tarapac\'{a}, Arica, Chile, under project 4721/2003 is gratefully
acknowledged by E. Lazo and M. E. Onell.

\newpage

\noindent

\end{document}